\documentclass[pre,preprint,showpacs]{revtex4-1}

\usepackage{amsmath}
\usepackage{mathrsfs}

\def\dif{\mathrm d}
\def \Dif {\mathcal{D}}

\newcount\minute
\newcount\hour
\newcount\hourMins
\def\now
{
  \minute=\time
  \hour=\time \divide \hour by 60
  \hourMins=\hour \multiply\hourMins by 60
  \advance\minute by -\hourMins
  \zeroPadTwo{\the\hour}:\zeroPadTwo{\the\minute}
}

\def\today
{
  \zeroPadTwo{\the\day}-\zeroPadTwo{\the\month}-\the\year%
}
\def\zeroPadTwo#1%
{
  \ifnum #1<10 0\fi
  #1%
}

\begin{document}

\title{Exact asymptotic behavior of correlation functions for disordered spin-1/2 XXZ chains}

\author{Zoran Ristivojevic,$^1$ Aleksandra Petkovi\'{c},$^1$ and Thierry Giamarchi$^{2}$}
\affiliation{$^1$Laboratoire de Physique Th\'{e}orique--CNRS, Ecole Normale Sup\'{e}rieure, 24 rue Lhomond, 75005 Paris, France}
\affiliation{$^{2}$DPMC-MaNEP, University of Geneva, 24 Quai Ernest-Ansermet, CH-1211 Geneva, Switzerland}

\date{\today \now}

\begin{abstract}
We consider an XXZ spin-1/2 chain in the presence of several types of disorder that do not break the XY symmetry of the system. We calculate the complete asymptotic form of the spin-correlation functions at zero temperature at the transition between liquid and disordered phase that occurs for a special value of anisotropy in the limit of small disorder. Apart from a universal power law decay of correlations, we find additional logarithmic corrections due to marginally irrelevant operator of disorder.

\end{abstract}


\maketitle

\section{Introduction}

A spin-1/2 XXZ spin chain is an exactly solvable model \cite{Baxter} and has the Hamiltonian
\begin{align}\label{H0}
H_{\scriptscriptstyle \mathrm{ XXZ}}=J\sum_i\left[(S^x_{i}S^x_{i+1}+S^y_{i}S^y_{i+1})
+\Delta S^z_{i}S^z_{i+1}\right],
\end{align}
where the spin can be taken as $\mathbf{S}=\boldsymbol{\sigma}/2$, where $\boldsymbol{\sigma}$ are the Pauli matrices. They satisfy the commutation relation $[S^a_j,S^b_k]=i\epsilon_{abc}S^c_j\delta_{jk}$. XXZ model exhibits rich physics when varying the anisotropy parameter $\Delta$. For $\Delta\le -1$ the ground state has Ising-type long-range ferromagnetic order and finite excitation gap in the low-energy spectrum. For $\Delta>1$ the ground state is again gapped and posses antiferromagnetic order. For $-1<\Delta\le 1$ the model is gapless and has a linear spectrum of low-energy excitations. For $-1<\Delta<1$, the staggered part\footnote{In the following we only consider the staggered part of correlation functions.} of its asymptotic correlation functions read \cite{Baxter,luther+75PhysRevB.12.3908}
\begin{align}\label{Gx}
&\langle S^x_0S^x_j\rangle\sim (-1)^j |j|^{-1/2K},\\
\label{Gz}
&\langle S^z_0S^z_j\rangle\sim (-1)^j |j|^{-2K},
\end{align}
where the parameter $K$ is given as
\begin{align}\label{DeltaKrelation}
\Delta=-\cos\frac{\pi}{2K}\quad (1/2\le K<\infty).
\end{align}
For $\Delta=1$ the model (\ref{H0}) becomes isotropic and thus the correlation functions (\ref{Gx}) and (\ref{Gz}) become equal. However they acquire an additional logarithmic factor \cite{affleck+89JPhysA22,giamarchi+89PhysRevB.39.4620,giamarchi+88JdePhys,singh+89PhysRevB.39.2562}
\begin{align}\label{GxGzHeisenberg}
&\langle S^x_0S^x_j\rangle=\langle S^z_0S^z_j\rangle\sim (-1)^j\frac{\log^{1/2}|j|}{|j|}.
\end{align}
Its origin is well understood. The low-energy properties of (\ref{H0}) correspond to a  $(1+1)$-dimensional sine-Gordon (SG) field theory. For $\Delta=1$ the corresponding SG model is exactly at criticality where its cosine term becomes marginally irrelevant and produces the above logarithmic corrections. Numerical works \cite{hallberg+95PhysRevB.52.R719,hikihara+98PhysRevB.58.R583} have confirmed its presence in (\ref{GxGzHeisenberg}).

An insight into log-corrections of the Heisenberg chain (XXZ for $\Delta=1$) comes from the fact that the SG model is equivalent to the classical two-dimensional XY model that exhibits a Berezinskii-Kosterlitz-Thouless (BKT) transition \cite{Berezinskii72,Kosterlitz+73}. In his early work \cite{Kosterlitz74JPhysC7}, Kosterlitz showed that the spin-spin correlation function at the transition contains logarithmic corrections of the form $\log^{1/8} |j|/|j|^{1/4}$ (there the spins are classical two-dimensional vectors). The ratio of powers of $\log |j|$  and $|j|$ is $1/2$, as in (\ref{GxGzHeisenberg}). That is easily understood since the power law exponent fixes the exponent of the logarithmic part for the same type of correlation functions, as in the case for the two examples. Having in mind the work \cite{Kosterlitz74JPhysC7}, it is surprising that the logarithmic corrections for correlations of the Heisenberg chain were rediscovered relatively late.

In the present paper we consider several types of disordered XXZ chains. The form of disorder is such that is preserves the XY symmetry of the pure model (\ref{H0}), see below. It turns out that all the disordered models we consider here have a BKT transition at $\Delta=-1/2$ in the limit of vanishing disorder strength \cite{doty+92PhysRevB.45.2167,Giamarchi+88}, and the region $-1<\Delta<-1/2$ (where the disorder is irrelevant) is a remnant of the region $-1<\Delta\le 1$ of the pure model, and therefore has quasi-long-range order and correlations (\ref{Gx}) and (\ref{Gz}), for weak disorder \cite{doty+92PhysRevB.45.2167}. We will calculate the spin-correlation functions at the transition. As in the pure case, they acquire again multiplicative logarithmic corrections, however the universal powers of logarithms are  now different for (\ref{Gx}) and (\ref{Gz}). That comes from the fact that the corresponding field theory is given by a disordered SG model.

The article is organized as follows. In Section II we introduce the different types of disorder and describe the low-energy properties of the system in terms of a 1+1-dimensional field theory. In Section III we compute the correlation functions of interest. Section IV contains conclusions. In Appendix we give the derivation of the scaling equations for the model.

\section{Disordered Hamiltonian and low-energy description}

\subsection{Different types of disorder}

We are interested in different types of disorder that preserve $XY$ symmetry of (\ref{H0}): (i) a random transverse magnetic field along $z$ direction,
\begin{align}\label{Hzf}
H_{\scriptscriptstyle \mathrm{ZF}}=\sum_i h_i^z S_i^z,
\end{align}
(ii) random $z-$exchange,
\begin{align}\label{Hze}
H_{\scriptscriptstyle \mathrm{ZE}}=\sum_i \delta J_i^z S_i^z S_{i+1}^z,
\end{align}
and (iii) random planar exchange interaction,
\begin{align}\label{Hpe}
H_{\scriptscriptstyle {\mathrm{PE}}}=\sum_i\delta J_i^{xy}(S^x_{i}S^x_{i+1}+S^y_{i}S^y_{i+1}).
\end{align}
Note that $h_i^z$, $\delta J_i^z$ and $\delta J_i^{xy}$ are random variables uncorrelated from site to site, each of them having the zero mean and the second moment determined by $D$:
\begin{align}
\overline{h^z_i h^z_j}=\overline{\delta J^z_i \delta J^z_j}=\overline{\delta J^{xy}_i \delta J^{xy}_j}=D\delta_{ij}.
\end{align}
The overbar denotes an average over different disorder configurations. In the following we consider the Hamiltonian
\begin{align}
H=H_{\scriptscriptstyle \mathrm XXZ}+H_{\scriptscriptstyle {\mathrm{D}}},
\end{align}
where $H_{\scriptscriptstyle {\mathrm{D}}}$ denote any one of the terms (\ref{Hzf}), (\ref{Hze}), or (\ref{Hpe}).

\subsection{XXZ chain}

First we start by considering the XXZ chain, which treatment is now standard \cite{sachdev,Giamarchi}. One translates the discrete model (\ref{H0}) into a continuous field theory that captures long-distance physics. Using the Jordan-Wigner transformation one first translates $\Delta=0$ part of (\ref{H0}) into noninteracting fermions. The remaining $z-$part essentially represents the interaction between fermions, since the spin operator $S^z$ becomes the density of fermions. After employing the bosonization representation one ends up with the continuum form of the action \cite{Giamarchi,sachdev}:
\begin{align}
&{\mathcal{S}_{\scriptscriptstyle {\mathrm{XXZ}}}}=\mathcal{S}_0+\mathcal{S}_g,\\\label{S0}
&\mathcal{S}_0=\frac{1}{2\pi} \int
\dif x\dif \tau\left[vK(\partial_x\theta)^2 +\frac{v}{K}(\partial_x\varphi)^2+2i(\partial_x\theta)(\partial_\tau\varphi)\right],\\
\label{Sg}
&\mathcal{S}_g=-g\int
\dif x\dif \tau\cos{(4\varphi)},
\end{align}
where $g\sim J\Delta$, and the spin-wave velocity is
\begin{align}
v=J a \frac{K}{2K-1}\sin{\left(\pi \frac{2K-1}{2K}\right)},
\end{align}
while the Luttinger parameter $K$ is connected to the anisotropy via (\ref{DeltaKrelation}). The lattice constant is denoted by $a$. The spin operators are connected to the bosonic fields $\varphi(x,\tau),\theta(x,\tau)$ through the following relations \cite{sachdev,Giamarchi}:
\begin{align}\label{sigma+-}
S^+_j&=(-1)^j\sum_{m \;\mathrm{even}}B_m \mathrm{e}^{im k_F x_j+i m\varphi(x_j)-i\theta(x_j)},\\
\label{sigmaz}
\frac{S^z_j}{a}&=-\frac{1}{\pi}\nabla \varphi(x_j)+\sum_{m\neq 0, \mathrm{even}}C_m \mathrm{e}^{im k_F x_j+i m \varphi(x_j)}.
\end{align}
where $x_j=j a$ and $k_F=\pi/(2a)$. Also, $B_m$ and $C_m$ are some nonuniversal constants.

\subsection{Disordered chain and scaling equations}

Next we include disorder. We start with the random field case (\ref{Hzf}) first. The corresponding action is
\begin{align}\label{Szf}
\mathcal{S}_{\scriptscriptstyle {\mathrm{ZF}}}=&\int \dif x\dif\tau \left\{-\frac{1}{\pi}\eta(x)\partial_x\varphi+2C_2\xi(x) \cos{(2\varphi)}\right\}+\ldots
\end{align}
where
\begin{align}
&\overline{\eta(x)\eta(x')}=D \delta(x-x'),\\ &\overline{\xi(x)\xi(x')}= D \delta(x-x').
\end{align}
Here the fields $\eta(x)$ and $\xi(x)$ are proportional to the uniform and alternating parts of the random field $h^z$. Higher order terms denoted by \ldots in (\ref{Szf}) correspond to less relevant higher harmonics from (\ref{sigmaz}). After performing the disorder average using the replica method, we find the action that corresponds to $H_{\scriptscriptstyle {\mathrm{XXZ}}}+H_{\scriptscriptstyle {\mathrm{ZF}}}$:
\begin{align}\label{S}
\mathcal{S}=\sum_{\alpha} \mathcal{S}_0(\varphi_{\alpha},\theta_{\alpha})+\mathcal{S}_f+\mathcal{S}_b.
\end{align}
The first term is given by (\ref{S0}). The off--diagonal quadratic part that comes due to disorder is
\begin{align}\label{Sf-rep}
\mathcal{S}_f=-\frac{1}{2\pi^2}D_f \sum_{\alpha \beta}\int
\dif x\dif\tau\dif\tau' [\partial_x\varphi_\alpha(x,\tau)][\partial_x\varphi_\beta(x,\tau')],
\end{align}
while the anharmonic part of the action reads as
\begin{align}\label{Sb-rep}
\mathcal{S}_b=-C_2^2 D_b \sum_{\alpha\beta}\int\dif x\dif\tau\dif\tau' \cos[2\varphi_\alpha(x,\tau)-2\varphi_\beta(x,\tau')].
\end{align}
Greek letters $\alpha,\beta=1,\ldots, n$ denote replica indices, where the number of replicas $n$ is sent to zero at the end of calculations. Note that only the most relevant terms are written, since they determine the critical behavior. The term (\ref{Sg}) turns out to be unimportant, since it is irrelevant close to the transition \cite{doty+92PhysRevB.45.2167}, occurring for $K=3/2$ ($\Delta=-1/2$). The term analogous to (\ref{Sb-rep}) but having plus sign in the argument of cosine is also suppressed in our calculation. The bare parameters $D_b$ and $D_f$ are the same and equal to $D$ in the unrenormalized model. However, they renormalize differently, hence we distinguish them.

The model (\ref{S}) is well known and describes disordered bosons and fermions in one dimension \cite{Giamarchi+88}, and via Jordan-Wigner transformation straightforwardly extends to disordered spin chains \cite{doty+92PhysRevB.45.2167}. It has the following scaling renormalization group equations \cite{Giamarchi+88}:
\begin{align}\label{D}
&\frac{\dif\mathscr{D}_R}{\dif\ell}=-2\mathscr{D}_R\delta_R +\mathcal{O}(\mathscr{D}_R^2),\\
\label{delta}
&\frac{\dif\delta_R}{\dif\ell}=-9\mathscr{D}_R +\mathcal{O}(\mathscr{D}_R\delta_R),\\
\label{Df}
&\frac{\dif }{\dif\ell}(aD_{fR})=0,\quad \frac{\dif}{\dif\ell}\left(\frac{v_R}{K_R}\right)=0,
\end{align}
where $\ell$ denotes the scale and the subscript index $_R$ denotes renormalized coupling constants that flow and that uniquely correspond to the bare ones. The dimensionless disorder strength is introduced as $\mathscr{D}=\pi C_2^2 a^3D_b/v^2$, while $\delta=K-3/2$ denotes the distance from the transition, occurring at $K=3/2$ or equivalently $\Delta=-1/2$, see (\ref{DeltaKrelation}). For $\Delta<-1/2$ weak disorder is irrelevant due to quantum fluctuations and at large scales one expects to have the properties of the XXZ chain. The phase for $\Delta>-1/2$ is disorder-dominated. Due to perturbative nature of the scaling equations, the above picture is valid for sufficiently weak disorder. Strong disorder even for $\Delta<-1/2$ is expected to dominate the physics.

The solution of the flow equations (\ref{D}) and (\ref{delta}) is
\begin{align}\label{solution}
\delta_R^2-9\mathscr{D}_R=C,
\end{align}
where $C$ is an arbitrary constant, and $C<0$ ($C>0$) marks the disorder dominated (quasi-long-range ordered XXZ) phase. For $C=0$ the system is at the transition, where the solution of (\ref{delta}) is
\begin{align}\label{flowdeltal}
&\delta_R(\ell)=\frac{\delta}{1+\ell\delta}.
\end{align}

\section{Correlation functions}

In this section we calculate the spin-correlation functions $\overline{\langle S^+_iS^-_{i+j}\rangle}$ and $\overline{\langle S^z_iS^z_{i+j}\rangle}$ at the transition between liquid and disordered phase, occurring for $C=0$ in (\ref{solution}). We focus below on their staggered part and moreover only on the most relevant contribution. We briefly discuss the uniform part of correlations at the end.

Using (\ref{sigma+-}), and (\ref{sigmaz}) we have:
\begin{align}\label{Spmdef}
\overline{\langle S^+_iS^-_{i+j}\rangle}&\sim (-1)^j \overline{\left\langle \mathrm{e}^{i\left[\theta(x_{i+j},0)-\theta(x_{i},0)\right]} \right\rangle}\equiv(-1)^j R_{+-}(ja,0),\\
\label{Szzdef}
\overline{\langle S^z_iS^z_{i+j}\rangle}&\sim (-1)^j a^2C_2^2\overline{\left\langle \mathrm{e}^{i2\left[\varphi(x_{i+j},0)-\varphi(x_{i},0)\right]}+\mathrm{h.c.} \right\rangle}\notag\\&\equiv(-1)^j a^2 C_2^2 \left[R_{zz}(j a,0)+\mathrm{h.c.}\right].
\end{align}
The average is with respect to the full action $\mathcal{S}_0+\mathcal{S}_{\scriptscriptstyle {\mathrm{ZF}}}$, Eqs.~(\ref{S0}) and (\ref{Szf}).

\subsection{$R_{zz}(x,\tau=0)$ correlation function}

The Callan-Symanzik equation determines the critical correlation function as \cite{Amit}
\begin{align}\label{eq:procedure}
R_{ab}(r,0;\delta,\mathscr{D})=\mathrm{e}^{\int_0^{\ell_r} \dif\ell\gamma_{ab}(\ell)}R_{ab}(a,0;\delta_R(\ell_r),\mathscr{D}_R(\ell_r)),
\end{align}
where $\ell_r=\ln(r/a)\gg 1$. Here
\begin{align}
\gamma_{ab}=\frac{\partial\ln Z_{ab}}{\partial \ln a},
\end{align}
where $Z_{ab}$ is the prefactor that multiplicatively removes the $a$-dependent divergences in $R_{ab}$. At large scales  $R_{ab}(a,0;\delta_R(\ell_r),\mathscr{D}_R(\ell_r))$  becomes a constant since the parameters $\delta_R(\ell_r),\mathscr{D}_R(\ell_r)$ reach their fixed point vales, that is zero in our case. Note that everywhere below in the text we use simplified notation $R_{ab}(r,0)=R_{ab}(r,0;\delta,\mathscr{D})$.

We first calculate $R_{zz}(r,\tau=0)$ ($r>0$) using the perturbation theory in the disorder strength \footnote{We warn the reader that the perturbation theory for correlation function is well defined in the limit $m\to 0$, while for the effective action calculated in Appendix we keep $m$ to be small and finite.}. Using the replica action we do the disorder average and should evaluate
\begin{align}
R_{zz}(r,0)=\lim_{n\to 0}\langle\mathrm{e}^{ i2\left[\varphi_\gamma(r,0)-\varphi_\gamma(0,0)\right]}\rangle_\mathcal{S},
\end{align}
where $\mathcal{S}$ is given by (\ref{S}). After evaluating Gaussian integrals one obtains
\begin{align}\label{R1bare}
&R_{zz}(r,0)=\mathrm{e}^{-\frac{2 K^2 D_f}{v^2}r}\left(\frac{a}{r}\right)^{3+2\delta} \left[1+W_1(r)+\mathcal{O}(\mathscr{D}\delta)\right],
\end{align}
where $W_1$ is
\begin{align}
\label{W1-calculated-special}
W_1(r)=&\frac{C_2^2 a^3 D_b}{v^2} \int\frac{\dif x\dif y\dif y'}{\left[a^2+(y-y')^2\right]^{3/2}} \left\{\left[\frac{A_1+a^2} {A_2+a^2} \frac{A_4+a^2} {A_3+a^2}\right]^{3/2}-1\right\}\notag\\
=&\frac{9}{4}\mathscr{D}\ln^2\frac{r^2}{a^2} +c_2\mathscr{D}\ln\frac{r^2}{a^2},
\end{align}
and $c_2$ being some number. We have introduced the abbreviations
\begin{align}\label{Aconstants}
&A_1=x^2+y^2,\quad A_3=(x-r)^2+y^2,\notag\\
&A_2=x^2+y'^2,\quad A_4=(x-r)^2+y'^2.
\end{align}
$W_1$ contains logarithmic divergencies for $a\to 0$. In order to calculate the long-distance behavior of $R_{zz}$ we should replace the bare parameters by the renormalized ones, which are connected via the relations:
\begin{align} \label{delta-deltaR} &\delta=\delta_R+9(c_1-\lambda)\mathscr{D}_R/2+\mathcal{O}(\mathscr{D}_R\delta_R),\\
\label{D-DR}
&\mathscr{D}=\mathscr{D}_R+\mathcal{O}(\mathscr{D}_R\delta_R).
\end{align}
Here $\lambda=\log{\left( c^2 m^2 a^2\right)}$ and $c$ is the constant. The details of calculations are presented in Appendix, see Eqs.~(\ref{Zb}) and (\ref{Z}). Using (\ref{delta-deltaR}) and (\ref{D-DR}) one ends up with
\begin{align}\label{R1renormalized}
R_{zz}(r,0)=&\mathrm{e}^{-\frac{2 K^2 D_{f}}{v^2}r}\left(\frac{r}{a}\right)^{-3-2\delta_R} \left[1-\frac{9}{4}\mathscr{D}_R\lambda^2 +\frac{1}{2}\left(9c_1-2c_2\right) \mathscr{D}_R\lambda\right]\notag\\
&\times\left\{1+\mathscr{D}_R\left[(2c_2-9c_1)\ln (cm r)+9\ln^2(cm r)\right] +\mathcal{O}(\mathscr{D}_R\delta_R)\right\},
\end{align}
so $R_{zz}$ is renormalizable by the multiplicative factor
\begin{align}\label{eq:Zzz}
Z_{R_{zz}}=(ma)^{-3-2\delta_R} \left[1+\frac{9}{4}\mathscr{D}_R\lambda^2 -\frac{1}{2}\left(9c_1-2c_2\right) \mathscr{D}_R\lambda\right].
\end{align}
Importantly all $r$-dependent divergencies for $a\to 0$ have canceled once one uses the renormalized parameters. Then one finds the anomalous dimension function
\begin{align}\label{eq:gammazz}
\gamma_{zz}=\frac{\partial \ln Z_{R_{zz}}}{\partial\ln a}=-3-2\delta_R+\mathcal{O}(\mathscr{D}_R,\delta_R^2).
\end{align}
Finally, using Eqs.~(\ref{flowdeltal}), (\ref{eq:procedure}), and (\ref{eq:gammazz}) we find the large scale behavior
\begin{align}\label{Rzzfinal}
  R_{zz}(r,0)\sim \mathrm{e}^{-\frac{2 K^2 D_f}{v^2}r}\left(\frac{a}{r}\right)^{3}\ln^{-2}\left(\frac{r}{a}\right),
\end{align}
where we used the fact that $D_f$ and $K/v$ are not renormalized, see (\ref{Df}).
We have found a logarithmic correction to the correlation function at  criticality. Note that higher order terms in (\ref{eq:gammazz}) do not give important contribution since after integration in (\ref{eq:procedure}) they saturate into a constant for large $x$. This can be seen as a change of the number under the logarithm that we neglected. The same type of contribution comes from  (\ref{flowdeltal}) at short length scales, while the universal large scale part $1/\ell$ determines the logs we find in (\ref{Rzzfinal}) and below in (\ref{Rpmfinal}).


\subsection{$R_{+-}(r,\tau=0)$ correlation function}

Next we calculate the other correlation function
\begin{align}
R_{+-}(r,0)=\lim_{n\to 0}\langle\mathrm{e}^{ i\left[\theta_\gamma(r,0)-\theta_\gamma(0,0)\right]}\rangle_\mathcal{S}.
\end{align}
Since it contains the conjugate field to $\varphi$, the disorder term proportional to $D_f$ does not appear in any order of perturbation theory in $R_{+-}$. One way to see that is to perform a gauge transformation by absorbing $D_f$ term in the new $\varphi$ fields. Under such transformation (\ref{Sb-rep}) is invariant. After performing the perturbation theory one obtains
\begin{align}
R_{+-}(r,0)=\left(\frac{a}{r}\right)^{\frac{1}{3+2\delta}}\left[1+W^\theta_1(r) +\mathcal{O}(\mathscr{D}\delta)\right]
\end{align}
where
\begin{align}
W^\theta_1(r)=&\frac{C_2^2 a^3 D_b}{v^3}\int\frac{\dif x\dif y\dif y'}{\left[a^2+(y-y')^2\right]^{3/2}}
\Bigg\{\left[\sqrt{\frac{A_1 A_4}{A_2 A_3}}-\frac{1}{2}\frac{(y-y')^2 r^2} {\sqrt{A_1 A_2 A_3 A_4}}\right]-1\Bigg\}\notag\\
=&-\frac{1}{4}\mathscr{D}\ln^2\frac{r^2}{a^2} +c_3 \mathscr{D}\ln\frac{r^2}{a^2}.
\end{align}
Here $c_3$ is some constant.
Note that $A_1,\ldots A_4$ have been defined in (\ref{Aconstants}).
Then one gets
\begin{align}\label{R2bare}
R_{+-}(r,0)=\left(\frac{a}{r}\right)^{\frac{1}{3+2\delta}}
\left[1-\frac{1}{4}\mathscr{D} \ln^2\frac{r^2}{a^2} +c_3\mathscr{D}\ln\frac{r^2}{a^2} +\mathcal{O}\left(\mathscr{D}\delta\right)\right].
\end{align}
Plugging in the renormalized parameters (\ref{delta-deltaR}) and (\ref{D-DR}) one gets
\begin{align}\label{R2renormalized}
R_2(r)=&\left(\frac{r}{a}\right)^{-\frac{1}{3+2\delta_R}}
\left[1+\frac{1}{4}\mathscr{D}_R \lambda^2 -\frac{1}{2}(2c_3+c_1) \mathscr{D}_R\lambda \right]\notag\\ &\times\left[1-\mathscr{D}_R\ln^2(cm r)+(2c_3+c_1) \mathscr{D}_R\ln(cm r) +\mathcal{O}(\mathscr{D}_R\delta_R) \right].
\end{align}
One should notice that all $r$-dependent divergencies in $a\to 0$ limit of the perturbation theory in (\ref{R2bare}) cancel once one expresses the bare coupling constants by the renormalized ones,  see Eq.~(\ref{R2renormalized}). We have the combination of the $\ln^2 (r/a)$ term from the right hand side of (\ref{R2bare}) that combines with $-\mathscr{D}_R\lambda\ln(r/a)$ that arises after expanding the power in (\ref{R2bare}) with the exponent $1/(3+2\delta)$ after using the renormalized quantities.

The previous expression (\ref{R2renormalized}) contains divergence when $a\to 0$ that could be removed multiplicatively by
\begin{align}
Z_{R_{+-}}=(ma)^{-\frac{1}{3+2\delta_R}}\left[1-\frac{1}{4}\mathscr{D}_R\lambda^2 +\frac{1}{2}(2c_3+c_1) \mathscr{D}_R\lambda \right],
\end{align}
so that the renormalized correlation function $Z_{R_{+-}} R_{+-}(r,0)$ has no dependence of $a$. The anomalous dimension function then reads
\begin{align}\label{eq:gammapm}
\gamma_{+-}=\frac{\partial \ln Z_{R_{+-}}}{\partial\ln a}=-\frac{1}{3}+\frac{2}{9}\delta_R+\mathcal{O}(\mathscr{D}_R,\delta_R^2),
\end{align}
where we have used the scaling equations (\ref{D}), and (\ref{delta}). Then, using Eqs.~(\ref{flowdeltal}),  (\ref{eq:procedure}), and (\ref{eq:gammapm}) we find the large scale behavior
\begin{align}\label{Rpmfinal}
  R_{+-}(r,0)\sim \left(\frac{a}{r}\right)^{1/3}\ln^{2/9}\left(\frac{r}{a}\right).
\end{align}

From (\ref{Spmdef}), (\ref{Szzdef}), (\ref{Rzzfinal}), and (\ref{Rpmfinal}) we obtain the equal time spin-spin correlation functions for $H_{\scriptscriptstyle \mathrm{ XXZ}}+H_{\scriptscriptstyle \mathrm{ZF}}$ at the transition (near $\Delta=-1/2$ and for weak disorder):
\begin{align}\label{Sxfinal}
&\overline{\langle S^+_iS^-_{i+j}\rangle}=2\overline{\langle S^x_iS^x_{i+j}\rangle}\sim\frac{ (-1)^j\ln^{2/9}|j|}{|j|^{1/3}},\\
\label{Szfinal}
&\overline{\langle S^z_iS^z_{i+j}\rangle}\sim \frac{(-1)^j}{|j|^3 \ln^2 |j|}\mathrm{e}^{-\frac{32D}{3J^2 a}|j|},
\end{align}
using the exact value at the transition $v/K=\sqrt{3} J a/4$.
We notice that the uniform non-staggered part \cite{sachdev} for $\overline{\langle S^+_iS^-_{i+j}\rangle}$ is subleading at large distances ($\sim |j|^{-10/3}$) therefore negligible with respect to the staggered part (\ref{Sxfinal}). However, the uniform part for $\overline{\langle S^z_iS^z_{i+j}\rangle}$ is $\sim |j|^{-2}$ and therefore dominates at large distances. Finally we notice that inside the logs of the last two formulas in general there are some nonuniversal disorder dependent numbers that come from small scales of the exponent in the Callan-Symanzik equation (\ref{eq:procedure}), that we neglected at large distances but that may be important for numerical analysis. The similarity of effective models for disordered spin chains and disordered bosons (and fermions) in one dimension directly determines the form for the single particle and density-density correlation functions for the latter, that respectively have a form similar to (\ref{Sxfinal}) and (\ref{Szfinal}) \cite{Ristivojevic+12}.

\subsection{Random exchange models}

Next we discuss disordered Hamiltonians given by (\ref{Hze}) and (\ref{Hpe}). At first glance they appear quite different from the random field case (\ref{Hzf}); however due to quantum fluctuations the critical phase (for $-1<\Delta<-1/2$ and small disorder) has the same properties for all three cases, (\ref{Hze}), (\ref{Hpe}), and (\ref{Hzf}) \cite{doty+92PhysRevB.45.2167}. That occurs because the disorder is irrelevant there. On the contrary, properties of disordered phases are different, since they are sensitive to the form of disorder. It turns out that critical properties at the transition from a phase with quasi-long-range order to a disorder-dominated phase are the same for all three cases \cite{doty+92PhysRevB.45.2167}. However, the corresponding replicated action for (\ref{Hze}) and (\ref{Hpe}) is given by $\mathcal{S}=\sum_{\alpha}\mathcal{S}_0(\varphi_{\alpha},\theta_\alpha)+\mathcal{S}_b$. We see  that the term (\ref{Sf-rep}), or equivalently the first term in (\ref{Szf}) of the unreplicated action, is missing for the two random exchange models \cite{doty+92PhysRevB.45.2167,Orignac+98PhysRevB.57.5812}. That has very important consequences for the $\langle S_i^z S_{i+j}^z\rangle$ correlation function, since it does not get suppressed by the disorder as in (\ref{Szfinal}). Therefore, one can repeat the above analysis and obtain the correlation functions at the transition. The first one will have the form as in (\ref{Sxfinal}), while the other one is now changed and reads
\begin{align}\label{Szfinalexchange}
&\overline{\langle S^z_iS^z_{i+j}\rangle}\sim \frac{(-1)^j}{|j|^3 \ln^2 |j|}.
\end{align}
The absence of disorder in (\ref{Szfinalexchange}) makes it universal at the transition, which one may detect in numerical simulations.

\section{Conclusions}

In this paper we have calculated exact asymptotic form of the spin-correlation functions for several types of disordered XXZ spin chains at the transition between the critical (disorder irrelevant) and disordered phases at zero temperature. We found the corresponding logarithmic corrections, given by (\ref{Sxfinal}) and (\ref{Szfinal}) for the random field case (\ref{Hzf}), and by (\ref{Sxfinal}) and (\ref{Szfinalexchange}) for the random exchange cases (\ref{Hze}) and (\ref{Hpe}). They appear due to marginally irrelevant operator of disorder, quite similar to the Heisenberg spin-chain case. While the prefactors in the correlation functions for the Heisenberg case are exactly known \cite{Affleck98,Lukyanov+97}, its precise determination in (\ref{Sxfinal})-(\ref{Szfinalexchange}) for the disordered case is a formidable task. However, if the disorder is weak one may expect the prefactors to have similar values as the ones of the pure XXZ case at $\Delta=-1/2$ that are exactly known \cite{Lukyanov99PhysRevB.59.11163}. Finally we mention that higher order scaling equations \cite{Ristivojevic+12} of the model (\ref{S}) would produce additive subleading terms in (\ref{Sxfinal})-(\ref{Szfinalexchange}), and therefore they are inessential. The presence of log-corrections at the transition may be important for numerical studies for precise determination of the phase boundaries, for example. They will also play a role in finite temperature quantities such as the spin susceptibilities and the NMR relaxation rate.

\acknowledgements
The work is supported by the ANR Grant No.~09-BLAN-0097-01/2 (A.P.~and Z.R.) and in part by the Swiss SNF under MaNEP and Division II (T.G.).

\appendix
\section{Effective action and scaling equations}

In this appendix we derive the scaling renormalization group equations for the model (\ref{S}). To treat the backward scattering part of the disorder, we use a field theoretic approach \cite{Zinn-Justin,Amit+80} to obtain the effective action of the model $\Gamma(\varphi)$. It reads \cite{Zinn-Justin}
\begin{align}\label{Gamma}
\mathrm{e}^{-\Gamma(\varphi)}=\int\Dif\chi \mathrm{e}^{-\mathcal{S}(\varphi+\chi)+\int\dif x\chi(x)\frac{\delta \Gamma}{\delta \varphi(x)}}.
\end{align}
This general equation can be solved perturbatively, order by order with respect to the small parameter $D_b$ \cite{Ristivojevic+12}. Up to an additive constant, the effective action reads
\begin{align}
\Gamma={\mathcal{S}_0+\mathcal{S}_f}+\frac{1}{\hbar}\langle \mathcal{S}_b(\varphi+\chi)\rangle_\chi+\mathcal{O}(D_b^2)
\end{align}
where $\langle\cdots\rangle_\chi$ denotes an average with respect to the field $\chi$ using the quadratic part $\mathcal{S}_0+\mathcal{S}_f$.

At the lowest order we easily obtain
\begin{align}\label{Gamma1}
\Gamma_1=&-\mathscr{B} \sum_{\alpha\beta}\int\dif x\dif\tau\dif\tau'\bigg\{\left[\mathrm{e}^{4G(0,\tau-\tau')}-1\right] \delta_{\alpha\beta}+1\bigg\} \cos[2\varphi_\alpha(x,\tau)-2\varphi_\beta(x,\tau')],
\end{align}
where $\mathscr{B}=C_2^2D_b\mathrm{e}^{-4G(0,0)}$, while
\begin{align}\label{G}
G(x,\tau)=\frac{K}{2}K_0(m\sqrt{x^2+v^2\tau^2+a^2}).
\end{align}
One should notice that only a part of the full correlation function that is diagonal in replica indices, (\ref{G}), enters the result. This is due to the fact that the off-diagonal part of the full propagator does not depend on imaginary time. We introduced $m$ as an infrared cutoff by adding a term $\propto m^2(\varphi_\alpha)^2$ into $\mathcal{S}_0$, in order to calculate the effective action.

The obtained perturbative expansion of the effective action contains all the information about critical properties of our model at the first order in $D_b$. In order to derive the scaling equations one should find the divergent terms in $\Gamma$ in the limit $a\to 0$. After expanding the operators one obtains the effective action

\begin{align}\label{Gammafinal}
\Gamma=&\sum_{\alpha} \int\dif x\dif\tau\left\{\frac{v}{2\pi K}\left[(\partial_x\varphi_\alpha)^2+m^2(\varphi_\alpha)^2\right]+ \left[\frac{1}{2\pi K v}+2\mathscr{B}a_1\right] (\partial_\tau\varphi_\alpha)^2\right\}\notag\\ &-\frac{D_f}{2\pi^2}
\sum_{\alpha\beta}\int\dif x\dif\tau\dif\tau'[\partial_x\varphi_\alpha(x,\tau)] [\partial_x\varphi_\beta(x,\tau')]\notag\\
&-\mathscr{B}\sum_{\alpha\beta}\int\dif x\dif\tau\dif\tau' \cos[2\varphi_\alpha(x,\tau)-2\varphi_\beta(x,\tau')].
\end{align}

The coefficient $a_1$ arises from the expansion of operators in (\ref{Gamma1}) and reads
$a_1=\int\dif\tau\tau^2\left[\mathrm{e}^{4G(0,\tau)}-1\right]$. Introducing the small parameter  $\delta=K-3/2$ which measures the distance from the critical point and using (\ref{G}) we get
\begin{align}\label{a1}
a_1=\frac{-\lambda+c_1+\mathcal{O}(\delta)}{(cmv)^3},
\end{align}
where $\lambda=\ln c^2m^2a^2$, $c=\mathrm{e}^{\gamma_E}/2$,  $\gamma_E$ is the Euler constant, and $c_1$ is a constant.

The effective action (\ref{Gammafinal}) has divergencies  when $ma\to0$ contained in $\lambda$. In order to remove them we introduce a set of renormalized coupling constants (denoted by the subscript $_R$) by $\mathscr{D}=Z_b \mathscr{D}_R,v=Z v_R,\delta=Z(3/2+\delta_R)-3/2,K=Z(3/2+\delta_R),m_R=m$
where
\begin{align}\label{Zb}
Z_b=&1-\delta_R\lambda,\\
\label{Z}
Z=&1-3\mathscr{D}_R\lambda+3c_1\mathscr{D}_R.
\end{align}
Differentiating the bare coupling constants with respect to the scale $\ell=-\ln m_R$ one obtains the scaling equations (\ref{D})--(\ref{Df}).

%

\end{document}